# Corporate Social Responsibility and Corporate Governance: A cognitive approach

*Rania Béji[a], Ouidad Yousfi[b] and Abdelwahed Omri[c]*

## Abstract

This chapter aims to critically review the existing literature on the relationship between corporate social responsibility (CSR) and corporate governance features. Drawn on management and corporate governance theories, we develop a theoretical model that makes explicit the links between board diversity, CSR committees' attributes, CSR and financial performance. Particularly, we show that focusing on the cognitive and demographic characteristics of board members could provide more insights on the link between corporate governance and CSR. We also highlight how the functioning and the composition of CSR committees, could be valuable to better understand the relationship between corporate governance and CSR.

**Keywords** Corporate Social Responsibility, Corporate Governance, Diversity, CSR committees, Corporate Social Responsibility Performance, Financial Performance

[a]*MRM, Université de Montpellier (France) and GEF2A, ISG de Tunis, Université de Tunis (Tunisia). Email :* r.beji@montpellier-bs.com*. Phone: 00 33 (0) 610349198*
[b]*MRM, Université de Montpellier (France). Email :* ouidad.yousfi@umontpellier.fr*. Phone : 00 33 (0) 499585169*
[c]*GEF2A, ISG de Tunis, Université de Tunis (Tunisia). Email :* abdelomri@gmail.com*. Phone : 00 216 71588443*

**Introduction**

Since the start of the new millennium, researchers have been looking for an alternative way to achieve sustainable development and better utilization of the firms' natural resources [Carvalho *et al*., 2018]. Companies have become more aware of the negative impacts of many business operations on the environment and society, specifically long-term effects. To overcome the short-term thinking, many firms have become more involved in socially and environmentally responsible strategies to improve their social, environmental and financial performances. This leads to the emergence of a more inclusive philosophy in businesses going beyond stockholders: CSR.

According to Ashrafi *et al*. [2018], CSR is a long-term approach that aims to bring a multi-dimensional added value to social, environmental, and economic spheres. Nowadays, responsible companies not only fully meet the applicable legal obligations, but also go beyond by extending their efforts to promote more socially responsible projects. For instance, companies, such as Starbucks and Orange, have become more concerned about the protection of human rights, employees' conditions, environmental issues, and communities' expectations.

According to Bocquet *et al*. [2017], Zerbini [2017], and Goyder [2003], there are two CSR strategies: (1) strategic CSR which ties in with the highest level of commitment and implies a more comprehensive implementation of CSR within a firm and (2) responsive CSR where CSR involvement is mainly determined by external expectations and reporting standards and corresponds to the lowest level of commitment.

CSR has been considered as one of the most important challenges of corporate governance. Companies and their boards of directors have to integrate socially responsible investment into their overall approach [Jamali *et al*., 2008]. Due to the increasing attention paid to CSR, scholars have examined the various antecedents of CSR.

On the one hand, there is an extensive literature on how CSR could influence the firm's performance and risks. Some CSR related studies have shown evidence that CSR activities lead to financial and non-financial benefits [Famiyeh, 2017; Hategan and Curea-Pitorac, 2017; and Reverte *et al*., 2016]. For instance, various empirical studies have concluded that socially responsible firms tend to have better social ratings and consequently are able to reduce their financial risk [Bouslah, *et al*., 2016; Harjoto and Jo, 2015; and Benlemlih *et al*., 2014]. Other studies have provided evidence on the existence of a negative correlation between CSR and financial performance [Galant and Cadez, 2017; Kim *et al*., 2014; Lee *et al*., 2013; and Baird *et al*., 2012]. We notice, however, that several emerging evidence shows insignificant relationships between the firm involvement in CSR activities and financial performance [Javed *et al*., 2016; Reverte *et al*., 2016; and Barnett and Salomon, 2012].

On the other hand, there is an emerging brand of the literature on the role of boards of directors in managing the firm's corporate image and shaping strategic orientations [Schepker *et al*., 2018; Schulze *et al*., 2001; Walsh and Seward, 1990]. In fact, CSR seems to be influenced by the choices, motivations, and values of those involved in the decision-making process. Accordingly, taking into account demographic attributes of directors such as gender, age, nationality, educational background, and multiple directorships, could be very helpful to better understand the board's dynamics and how they could influence the firm performances from different perspectives [Won-Yong *et al*., 2019; Tasheva and Hillman, 2018; Haniffa and Cooke, 2005; and Gibbins *et al*., 1990]. For instance, more diverse attributes on the board could lead to better organizational outcomes [Won-Yong *et al*., 2019]. Besides, the resource dependence [Pfeffer and Salancik, 1978], resource-based view [Wernerfelt, 1984; and Barney and Tyler, 1991], social categorization [Tajfel, 1981], upper echelons [Hambrick and Mason, 1984], and agency theories provide strong arguments on how more diverse boards could lead to superior monitoring and advisory capabilities, and therefore a more strategic involvement in CSR [Aggarwal *et al*., 2019; and Tasheva and Hillman, 2018].

On the same vein, there is an emerging literature on the board organization and the way boards conduct their roles and their influence on financial and CSR performances. More specifically, the role and the composition of specialized committees such as sustainable development, CSR, nomination, and compensation committees, seem to be a key determinant of CSR performance and CSR-related issues. For instance, Peters and Romi [2015] and Rodrigue *et al*. [2013] argue that CSR committees could be an essential part of the corporate governance structure. They aim to guide the company towards CSR actions, as well as to promote and implement firms' CSR initiatives, which decreases CSR risks and achieve new opportunities. Moreover, Khan [2017] argues that the existence of CSR committees acts as an effective mechanism to enhance CSR performance. For instance, it can provide new incentives to CEOs to promote CSR strategies and to fulfill, therefore, sustainable development goals. Similarly, according to Hussain *et al*. [2018a], CSR committees establish the rules required to promote CSR activities. They control the impacts of companies' activities that affect or are affected by their operations on different stakeholders' groups such as environmentalists, the community, employees, consumers, and suppliers.

The main aim of the current chapter is to cover these challenging issues based on a critical state of the art. In particular, this chapter stresses how board features could be a determinant key to CSR decision making processes. Also, it and provides new perspectives to improve our understanding of the role of diversity in boards in CSR performance. From a managerial angle, it helps to assess how the board strategies and human resources could drive the firms to achieve a better CSR performance and more competitive advantages [Galbreath, 2010; Porter and Reinhardt, 2007; and Hart, 1995].

This chapter is structured as follows. First, we analyze how board composition could influence the CSR and financial performances. We distinguish between diversity in boards and diversity of boards. We identify the main features of strategic CSR and responsive CSR activities. Section (2) focuses on the board functioning and brings some light on the impact of the committee's boards, specifically CSR committees, and their role on CSR strategies and CSR-related decisions. A critical analysis is provided in Section (3). The last section of the chapter provides concluding insights.

## (I) Corporate Social Responsibility: could Ethics rebuild Financial Performance?

### I.1 On the confounded link between social performance and financial performance

The debate over the relationship between corporate social performance (CSP) and corporate financial performance (CFP) has dominated the empirical research during the last 40 years [Wood, 2010; Preston and O'Bannon, 1997; and Wood, 1991].

The results of recent empirical studies remain very mixed [Rost and Ehrmann, 2017]. More than 200 empirical studies have been reviewed in previous research [Lu *et al*., 2014; Nelling *et al*., 2009; Margolis *et al*., 2007; Allouche and Laroche, 2005; Margolis and Walsh, 2003; and Orlitzky *et al*., 2003]. The majority of studies show the existence of a positive association between social performance and financial performance [see among others Nelling *et al*., 2009; Margolis *et al*., 2007; and Orlitzky *et al*., 2003].

In fact, three theoretical models have been developed. The first model describes a positive association between CSP and CFP. The results of several meta-analyses confirm the existence of a positive relationship [Rost and Ehrmann, 2017; Endrikat *et al*., 2014; Albertini, 2013; Dixon-Fowler *et al*., 2013; Margolis *et al*., 2009; and Orlitzky and Swanson, 2008]. This model is based essentially on the theory of social impact [Cornell and Shapiro, 1987] rooted in the stakeholder theory of Freeman [1984]. It is based on the assumption that a company creates a competitive advantage through the ability to acquire resources, which leads to the establishment of a long-term synergy between the CSP and the CFP [Barney, 1991]. Furthermore, previous studies point out that meeting the needs of the different stakeholders acts as an insurance tool covering reputation risk during crises [Peloza, 2006; Schnietz and Epstein, 2005; Ziglidopoulos, 2001; and Orlitzky and Benjamin, 2001]. Also, Lu *et al*. [2014] reported an inconclusive overall result of the CSP-

CFP link through a meta-analysis of 84 empirical studies from 2002 to 2011. They find a significant positive relationship between CSP reputation ratings and CFP. Several other studies rooted in the social impact theory and the slack resources theory, such as Waddock and Graves [1997], state that better CFP is a source of social performance. Although companies follow the normative rules of good corporate citizenship, their actual behavior may depend on the resources available. Accordingly, profitability can increase a company's ability to fund social performance projects.

The second model describes a negative link between CSP and CFP [Shane and Spicer, 1983; Freedman and Jaggi, 1982; and Vance, 1975].

Based on the trade-off hypothesis and the hypothesis of the opportunism of managers, CSR decreases financial performance.

The first hypothesis states that the investments in CSR activities may worsen a firm's profitability by inhibiting optimal resource allocation, thus creating a competitive disadvantage [Kang *et al*., 2010]. The second hypothesis assumes that corporate executives can pursue their private benefits at the expense of shareholders and stakeholders' interests [Weidenbaum and Vogt, 1987, Williamson, 1985, 1967]. One explanation could be that when financial performance is high, managers can seize the opportunity to increase their gains by reducing social spending [Bénabou and Tirole, 2010; and McWilliams *et al*., 2006].

The third model points out that the costs and benefits of CSR cancel each other out [McWilliams *et al*., 1999]. In fact, previous studies argue that the link between CSP and CFP may not exist [Germann *et al*., 2015; Guiral, 2012; Goll and Rasheed, 2004; McWilliams and Siegel, 2000; and Aupperle *et al*., 1985]. In fact, CSP-CFP relationship could be powered by many confounding variables, such as environmental and social regulations stringency, R&D expenses, advertising expenditures, labor market conditions, *etc*.

Besides, Allouche and Laroche [2005a] and Orlitzky *et al*. [2003] show the existence of a virtuous circle, where a financially successful company is prone to be able to spend more on social activities, and better CSP provides a superior long-term economic return.

CSR literature has begun to question the validity of previous studies on the relationship between CSP and CFP since many researchers have displayed discrepancies [Shahzad and Sharfmann, 2017; Crane *et al.*, 2017; Jean *et al.*, 2016; Endrikat *et al.*, 2014; Schreck, 2011; and Garcia-Castro *et al.*, 2010]. For instance, Crane *et al.* [2017] and Shahzad and Sharfmann [2017] point out that endogeneity could be the main reason for the ambiguity of the results, specifically in structural equation modeling (SEM) studies using regression analysis to extract causal inferences [Jean *et al.*, 2016]. This may change the direction and the amplitude of CSR-CFP relationship, and distort results' interpretation [Ketokivi and McIntoch, 2017; and Zaefarian *et al.*, 2017].

The mixed results on CSP-CFP relationship could be explained to a large extent by the difference in terms of CSR strategies adopted by firms. In fact, CSR practices vary significantly according to the firm's degree of involvement in CSR behavior. Some firms are involved in strategic CSR: they are pioneers and want to achieve a better CSP by going beyond mandatory rules and standards. Some other firms adopt responsive CSR strategies and get involved only on what is legally expected or compliant with stakeholders' demand.

Hereafter, we analyze the differences between the two strategies and how they could shape CFP and firms' strategies.

### I.2 Strategic CSR versus responsive CSR

Recently, CSR has been directly associated with firms' performance [see among others Vilanova *et al.*, 2009; and Porter and Kramer, 2006]. According to Amin-Chaudhry [2016] and Bagnoli and Watts [2003], firms engage in profit-maximizing CSR, being their lead motivation. Therefore, the proponents of CSR are convinced that investment in CSR enhances the firm's long-term revenue and reputation [Manzano and Fernandez, 2016; Burke and Logsdon, 1996]. In fact, Porter and Kramer [2006] point out that when firms integrate CSR activities into their practices, this would help them in achieving competitive advantages. In consequence, according to Zerbini [2017], Bocquet *et al.* [2017], Porter

and Kramer [2006], Goyder [2003], and Burke and Logsdon [1996], two opposite views of CSR emerge:

1-      Strategic CSR where CSR and the firm's core competencies and resources are aligned. According to Burke and Logsdon [1996], CSR becomes strategic when the company considers social and environmental issues as a high priority and goes beyond the implementation of best practices. It consists of aligning all considerate acts towards people and the natural environment in the organizational context beyond the legal minimum, with the overall objectives and actions of a company. Besides, Bansal *et al*. [2015] argue that the strategic CSR comprises activities with long time horizons and large resource commitments. It also allows the firm to achieve a distinctive position as compared to competitors [Burke and Logdson, 1981].

2-      Responsive CSR where CSR involvement is mainly determined by external expectations and reporting standards. The basic aim of firms is image-building to gain legitimacy in the eyes of their stakeholders [Ruggiero *et al*., 2018]. Porter and Kramer [2006] define responsive CSR as mitigating existing or potential adverse effects of organizational activities. Indeed, responsive CSR is, most often, associated with a limited level of commitment and more adaptive behavior.

In fact, several relevant theoretical perspectives provide an understanding of the strategic CSR [Bhattacharyya *et al*., 2008; McAlister and Ferrell, 2002; and Burke and Logsdon, 1996].

The first framework is the natural resource-based view proposed by Hart [1995]. Specifically, Hart [1995] classifies product stewardship, pollution prevention and sustainable development as three interconnected strategies to support a natural resource-based view. He also points out that achieving sustainable development is the most demanding. It requires good stakeholder integration and good planning. In fact, this framework argues that by integrating sustainability and technological innovation into their strategies, firms could acquire a competitive advantage.

The second framework is the strategic CSR proposed by Burke and Logsdon [1996]. In fact, the authors propose five strategy dimensions to differentiate strategic CSR from responsive CSR: (1)

centrality ("the closeness of fit to the firm's mission and objectives"); (2) proactivity ("the degree to which the program is planned in anticipation of emerging social trends and the absence of crisis"); (3) voluntarism ("the scope for discretionary decision-making and the lack of externally imposed compliance requirements"); (4) visibility ("observable, recognizable credit by internal and/or external stakeholders for the firm"); (5) specificity ("the ability to capture private benefits by the firm"). Burke and Logsdon [1996] point out when firms' CSR initiatives meet these features, they are more likely to generate economic benefits. Accordingly, strategic CSR could positively impact firm financial performance.

The third framework is provided by Porter and Kramer [2002, 2006, 2011]. They argue that strategic CSR goes beyond best practices and provides a competitive advantage, while responsive CSR concerns acting as a good corporate citizen and responding to stakeholders' demands. Accordingly, choosing strategic or responsive CSR produces varied benefits [Bocquet *et al*., 2019; Martinez-Conesa *et al*., 2017; Chang, 2015; and Bocquet *et al*., 2013]. In fact, if a company combines effectively its resources and expertise with the competitive context, CSR could drive an integral part of its profitability and its competitive positioning [Porter and Kramer, 2002, 2006, 2011]. The strategic CSR approach suggested by the authors aims to achieve convergence between social and economic objectives, by requiring firms to use their attributes to meet social needs. Specifically, this model encourages companies to be more selective in terms of CSR engagement.

The final theoretical framework is the stakeholder theory [Freeman *et al*., 2004]. In fact, managing complex stakeholder relationships is considered as one of the main reasons why a firm should be more concerned about CSR and the opportunities that stakeholders could bring to the business [Post, 2003]. The instrumental theory of stakeholders considers CSR as a strategic driver of wealth creation [Garriga and Melé, 2004; Jones, 1995]. Also, according to Jamali [2008] and Turker [2009], strategic CSR studies are aligned with the stakeholder perspective of CSR. This could serve as useful guidance for managers in their pursuit of CSR, by providing an easier explanation of stakeholder issues.

All of the above theories on the link between CSP and CFP, provide strong support for the positive association between corporate governance and CSR activities.

In fact, the academic debates surrounding this approach argue that the association between good CSR policy and the appropriate behavior of board directors could improve financial profitability [Kaufman and Englander, 2011; Choi *et al*., 2010; Pesqueux and Damak-Ayadi, 2005; Donaldson, 1999; Jones and Wicks, 1999; Preston and Donaldson, 1999; Sternberg, 1999; and Freeman, 1984]. For instance, according to Cuervo [2002], corporate governance is a specific mechanism where the board of directors plays a relevant role in advising management on taking the most appropriate decisions and ensuring the long-term viability of the company. Consequently, the decisions taken by the board could lead to the possible implementation of CSR policies [Ingley *et al*., 2011] and influence, therefore, CFP.

Moreover, Choi *et al*. [2010] argue that the level of effort in terms of CSR depends on the relative importance given by the company to its interest groups. The company establishes an order of priority amongst them and favors those who are best positioned [Surroca *et al*., 2010].

Thus, it becomes imperative to introduce good corporate governance recommendations as an important element of CSR.

In the next section, we analyze how governance features could be influential to different extents on the degree of involvement in CSR strategies. Specifically, we focus on board characteristics.

### (II) Corporate Governance and CSR

Research and scholarship on board diversity have been one of the most prolific topics in recent years, appearing as one of the most significant current themes in corporate governance research [Tasheva and Hillman, 2018; Jizi, 2017; Harjoto *et al*., 2015; Hafsi and Turgut, 2013; Ben-Amar *et al*., 2013; Mahadeo *et al*., 2012; Bear *et al*., 2010; Kang *et al*., 2007].

The prior literature has focused on the role of board diversity on cognitive impacts such as creativity, innovation and the generation of new ideas [Tasheva and Hillman, 2018; Adams *et al*., 2015; Miller and Triana, 2009; Kang *et al*., 2007; Ruigrok *et al*., 2007; Carter *et al*., 2003; and Robinson and Dechant, 1997]. These studies show that diversity fosters creativity, innovation, and independence of thought processes. For instance, Carter *et al*. [2003] argue that board diversity allows a company to better understand her marketplace, which improves market penetration ability.

According to several studies [Mahadeo *et al*., 2012; Kang, 2007; Erhardt *et al*., 2003; Milliken and Martin, 1996], board diversity refers to the heterogeneity of a board across different demographic characteristics such as age, educational level, gender, and nationality, as well as the heterogeneity across different structural characteristics such as duality and independence. For instance, Mahadeo *et al*. [2012] and Carter *et al*. [2003] argue that diversity can be considered as an ethical objective, which brings to the company several advantages, such as greater creativity, a better understanding of the market, and better problem-solving. Therefore, it provides a competitive advantage for the company, as well as several beneficial long-term results [Erhardt *et al*., 2003; Siciliano, 1996].

Furthermore, Hemmingway and Maclagan [2004] argue that individuals' beliefs and values could influence board discussion related to CSR, as there is no mandatory standard for CSR [Deegan *et al*., 2006]. Also, Hambrick *et al*. [1996] and Nielsen [2010] put forward that, in high uncertainty contexts, diverse teams are more successful, while, in stable contexts, less diverse teams achieve better performance.

In fact, many theories have highlighted the effective role of board members to implement effective CSR strategies. For instance, the upper echelons theory [Hambrick and Mason, 1984] provides strong arguments on how more diverse boards could lead to superior monitoring, and thus, more strategic involvement in CSR [Aggarwal *et al*., 2019; and Tasheva and Hillman, 2018]. This theory suggests that the characteristics of directors, such as age, gender, educational level, knowledge, skills, values, professional experience, and tenure, influence their interpretations of the situations they face, which affects, therefore, their strategic choices [Hambrick, 2007].

Based on agency theory [Jensen and Meckling, 1976], one of the main functions of a board of directors is to act as fiduciaries of shareholders by monitoring top management on behalf of shareholders. According to Jo and Harjoto [2011, 2012], the effectiveness of corporate governance practices is closely related to the board's composition. For instance, corporate transparency practices are determined by board directors to improve management practices and to ensure the involvement of a company in more ethical projects.

Furthermore, another important theoretical perspective for diversity is the resource dependence theory [Pfeffer and Salancik, 1978]. In fact, several studies point out that board diversity allows the firm to acquire critical resources, policy advice, knowledge, and networks, and to widen the range of choices when making strategic decisions [Locke and Reddy, 2015; Taljaard *et al*., 2015; Al-Musalli *et al*., 2012; Goodstein *et al*., 1994; Pfeffer, 1972; and Pfeffer and Salancik, 1978]. According to Bear *et al*. [2010], diversity enhances the internal and external resources of the board, such as the new skills and competencies. This helps companies to better respond to stakeholders' expectations. They become more sensitive to CSR issues due to the variety of resources given by board diversity [Davis and Cobb, 2010; Susan Vinnicombe *et al*., 2003; and Pfeffer and Salancik, 1978].

## II.1 Board diversities: diversity in boards and diversity of boards

Much of the prior literature has focused on the impact of board composition on a firm's behavior [Lehn *et al*., 2009; Linck *et al*., 2008; and Fich and Shivdasani, 2006]. The level of diversity could be considered as one of the key aspects of board composition.

Prior research on board diversity distinguishes between structural and demographic diversity of boards [Aggarwal *et al*., 2019; Jizi, 2017; Harjoto *et al*., 2015; and Hafsi and Turgut, 2013]. The structural diversity of boards is linked to dissimilarities in board attributes, which are related to boards' formal structure, such as size, the non-separation between management and control functions, board independence, and board committees [Tasheva and Hillman, 2018]. Demographic diversity, however, is given by the individual attributes of directors such as gender, age, nationality, educational level, educational background, multiple directorships, culture, tenure, nationality, and experience. Hafsi and Turgut [2013] highlight that the structural diversity of boards does not allow to differentiate among firms or to explain their differences, while demographic diversity does. Table 1.1 summarizes the findings from this literature review.

Table 1.1.a. Summary of studies on the impact of board diversity on CSR

| Author(s) | Aim | Method | Diversity measure | Sample and time | Key findings |
|---|---|---|---|---|---|
| Olthuis and van den Oever [2020] | The impact of the ideological diversity on CSR performance | Quantitative (regression) | Board Ideological diversity | Dutch firms (2014-2017) | A negative relationship between board ideological diversity and CSR performance. |
| Cordeiro et al. [2019] | The impact of ownership structure and board gender diversity on corporate environmental performance | Quantitative (regression) | Gender diversity Ownership structure | U.S. firms (2010-2015) | The majority of family owners and dual-class owners interact with board gender diversity to positively influence corporate environmental performance. |
| Harjoto et al. [2019] | The impact of the nationality and educational background diversity on CSR | Quantitative (regression) | Board nationality Educational background diversity | U.S. firms (2000-2013) | Improving director nationality diversity and educational background increase firms' social performance. |
| Galbreath [2018] | The impact of board gender diversity on financial and social performance | Quantitative (regression) | Gender diversity | Australian firms (2004-2005) | Gender diversity influences firms' prosocial actions, which results in higher levels of CSR. |
| Cuadrado-Ballesteros et al. [2017] | The impact of the board of directors on CSR practices. | Quantitative (regression) | Board size Board independence Diversity of gender and nationality | Firms from Canada, Denmark, Finland, France, Italy, the Netherlands, Norway, the UK, the USA, Germany, Spain, and Sweden. (2003-2009) | Board size is positively associated with CSR practices; however, when the number of directors is excessively high, the CSR commitment decreases. board diversity increases the level of CSR practices. |
| Ben Barka and Dardour [2015] | The impact of board interlocks, director's profile on CSR | Quantitative (regression) | Director's background Nationality diversity Age Gender Tenure Board size Ownership Duality Independence | France (2010) | Director's background and nationality diversity are the most relevant attributes to discerning firms with high CSR scores. |

Table 1.1.b. Summary of studies on the impact of board diversity on CSR

| Author(s) | Aim | Method | Diversity measure | Sample and time period | Key findings |
|---|---|---|---|---|---|
| Bouloua [2013] | The impact of female directors on CSR | Quantitative (regression) | Female directors | U.S. firms (1999-2003) | More gender-diverse boards exert a stronger influence on CSR performance. |
| Hafsi and Turgut [2013] | The impact of board diversity on CSP | Quantitative (regression) | Director ownership CEO duality Board independence Director tenure Director ethnicity Director age Director experience Gender diversity | U.S. firms (2005) | A significant relationship between demographic diversity and social performance, which is moderated by the structural diversity of boards. In particular, gender, and age have a significant effect on corporate social performance. |
| Post et al. [2011] | The impact of Board composition on environmental performance | Quantitative (regression) (disclosure-proxy) | Age Gender diversity Outside directors Cultural background Educational attainment | U.S. firms (2006-2007) | A higher proportion of outside board directors is associated with better environmental performance and higher KLD strengths scores. Firms with boards composed of three or more female directors have higher KLD strengths scores. And, boards whose directors average closer to 56 years are likely to implement environmental governance structures. |
| Jo and Harjoto [2011] | The impact of CG on the choice of CSR | Quantitative (regression) | Board independence Board leadership | U.S. firms (1993-2004) | CSR choice is positively associated with board leadership and board independence. |
| De Villiers et al. [2011] | The impact of board characteristics on environmental performance | Quantitative (regression) | Board diversity Board size Board independence Legal experts Active CEO CEO duality | U.S. firms (2003-2004) | Larger boards, higher board independence, lower duality, more legal experts on the board, and larger representation of active CEOs on the board increase environmental performance. |
| Bear et al. [2010] | The impact of diversity of board resources and the number of women on boards on CSR ratings | Quantitative (regression) | Gender diversity Director resource diversity: variety of experience and knowledge | U.S. firms and international firms (2009) | Gender diversity increases CSR ratings. |

*II.1.1. Diversity of boards*

Diversity of boards stems from the structural differences that could exist between boards. It has several attributes.

First, board size is considered as a critical factor that determines the effectiveness of board oversight. However, the literature provides no consensus regarding the effect of board size on firm performance. The resource dependency theory holds that large boards are likely to have better information and more knowledge, which allows them to provide more oriented advice on strategic decisions [De Villiers et al., 2011; Siciliano, 1996; Provan, 1980; and Pfeffer, 1972, 1973]. For instance, Kabir et al. [2017] put forward that large boards can increase the firm's involvement in CSR investments and have a better CSR performance as they have more resources provided for consulting and monitoring roles. Moreover, prior studies show that large boards are more likely to constitute a specific social capital and have more effective communication. Therefore, they are prone to contribute more to the efficiency of the ethical decision-making process than small boards, which could lead to a better CSR performance [Hillman et al., 2001; Clarkson 1995; and Pfeffer and Salancik, 1978].

For instance, the mean board size of a firm is around 9 members in European firms and US firms [Haque and Jones, 2020; and Madden et al., 2020].

Table 1.2 presents descriptive statistics of the study of Beji et al. [2020] on the association between boards' characteristics and globally CSR and specific areas of CSR, conducted on French companies listed on the SBF 120[d] index between 2003 and 2016. It uses Bloomberg, Factset IODS, and Thomson Reuters for financial and corporate governance data, and VigeoEiris[e] for CSR scores.

---

[d] The SBF120 index consists of the largest 120 capitalizations listed on the French Stock Exchange market (SBF: Société des Bourses Françaises).

[e] http://vigeo-eiris.com/fr/

Table 1.2 Descriptive statistics

| VARIABLE | N | MEAN | STD. | MIN | MAX |
|---|---|---|---|---|---|
| **BOARD SIZE** | 937 | 12.8943 | 3.5268 | 3 | 23 |
| **INDEPENDENCY** | 937 | 52.5293 | 21.4000 | 0 | 100 |
| **DUALITY** | 937 | .33617 | .4726 | 0 | 1 |
| **GENDER** | 937 | 22.1631 | 13.9675 | 0 | 63.6363 |
| **AGE** | 937 | .6247 | .1068 | 0 | .7901 |
| **FOREIGN** | 937 | 23.5345 | 21.1733 | 0 | 100 |
| **EDUCATIONAL-** | 937 | 69.6529 | 22.2949 | 0 | 100 |
| **BUSINESS-** | 937 | 63.2988 | 18.4547 | 14.2857 | 100 |
| **MULTIPLE-** | 937 | 73.30 | 16.7990 | 9 | 100 |

However, the agency theory holds that agency problems can become more severe with larger boards, specifically when they suffer from coordination and communication problems [Hermalin and Weisbach, 2003; Bushman and Smith, 2001; and Yermack, 1996]. In fact, it is easier for the CEO to control and influence the smaller boards as they can reach consensus more easily in comparison with large boards [Cheng, 2008].

Another board characteristic is the presence of independent directors on boards [Hermalin and Weisbach, 2003]. Harjoto and Jo [2011] point out that independent directors have stronger stakeholder orientation and better management quality, which leads to a successful CSR implementation [Shaukat *et al*., 2015; Li *et al*., 2012; Harjoto and Jo, 2011; Ho and Wong, 2001]. Also, Independent directors are prone to reduce agency conflicts and to ensure effective monitoring. In the same vein, Adams and Ferreira [2009] and Walsh and Seward [1990] point out that independent directors help to monitor executives' agency behavior, as they tend to check managers' self-serving decisions and to solve attendance problems on the board.

Also, previous studies argue that the duality structure on the board is prone to decrease corporate investment in CSR. According to Jizi *et al*. [2014] and Surroca and Tribo [2008], the concentration of management and control functions in the CEO's hands is likely to have negative impacts on the engagement in CSR activities. Furthermore, in line with agency theory, duality increases the CEO power and, therefore, CEO-chair could enjoy private benefits at the expense of CSR investments. In fact, CEOs who also act as the chair may pursue opportunistic strategies to protect their interests at the expense of shareholders [Jizi et al., 2014; and Firth et al., 2007]. They are also prone to prefer short term financial projects and to marginalize value-enhancing projects, specifically long-term projects such as CSR ones [Surroca and Tribo, 2008; and Firth et al., 2007].

Focusing only on the diversity of boards does not help to fully assess the most influential factors of CSP. We need also to address its interaction with other forms of diversity, such as diversity in boards.

## II.1.2 Diversity in boards

There is emerging literature, specifically on cognitive governance, on the influence of board members on CSP. It shows that, under specific conditions, the diversity of directors 'profiles could be a valuable resource for the business, particularly on CSR area [see among others Conyon and He, 2017; Rodriguez Ariza et al., 2016; Pucheta et al., 2016; and Boulouta, 2013].

Gender diversity on boards has attracted increasing interest in the last years. According to the social role theory [Eagly, 1987; and Eagly and Wood, 1991], women are more likely to be oriented toward others' welfare, more concerned with personal relationships, and more socially skilled than men. They are prone to show communal qualities, while men are likely to display agentic qualities [Eagly and Wood, 1991]. Moreover, in line with the cognitive moral reasoning theory (Kohlberg, 1969, 1976], women and men are different in terms of moral reasoning [Jaffee and Hyde, 2000]. In fact, previous studies show that women have higher cognitive moral reasoning scores and more ethical perceptions than men [Elm et al., 2001; Eynon et al., 1997; Forte, 2004].

Also, consistent with the upper echelons' theory [Hambrick and Mason, 1984], women and men on board display different cognitive features.

They have different norms, attitudes, perspectives, experiences, and knowledge [Sundarasen *et al*., 2016; and Pelled *et al*., 1999]. For instance, prior studies show that female directors could bring to light new perspectives, which improves, therefore, the governance quality [Conyon and He, 2017; Pucheta *et al*., 2016; and Krishnan and Parsons, 2008]. Furthermore, in line with the resource dependence theory [Pfeffer and Salancik, 1978] and the social identity theory [Ashforth and Mael, 1989], female directors are more engaged in social activities and more likely to undertake non-profit activities. Besides, they could provide new perspectives and many resources to the board which improves, therefore, the governance quality [Conyon and He, 2017; Rodriguez Ariza *et al*., 2016; Pucheta *et al*., 2016; Boulouta, 2013; Zhang, 2012; Post *et al*., 2011; Nielsen and Huse, 2010; and Krishnan and Parsons, 2008].

Previous studies put forward that socially responsible firms are associated with a higher percentage of female directors [Harjoto *et al*., 2015; Hafsi and Turgut, 2013; Zhang *et al*., 2012; and Carter *et al*., 2003]. For instance, Carter *et al*. [2003] find that gender-diverse boards perform better than less diverse ones. Further, Rodriguez Ariza *et al*. [2016], Braun [2010], and Nielsen and Huse [2010] point out that female directors are prone to be more engaged in green activities and more concerned about environmental issues than men. However, low gender quotas cannot influence CSP. For example, Post *et al*. [2011] show that environmental strengths scores increases in the presence of at least three female members on the board.

In fact, using a dataset of listed firms from India, China, and Russia over the period 2007-2014, Saeed and Sameer [2017] find that board gender diversity had increased in each country. Specifically, gender diversity on boards had increased from 5.6% to 10.3% in India, and 4.7% to 9.5% for Russia. The highest increase has been noticed in China where the percentage of female directors has more than doubled (from 7.6% to 14.4%). This could be explained by many reasons such as globalization and the proliferation of cross-border trade, communities' pressure (feminist groups), the emerging of standards and norms for acceptable governance practices conducted by international organizations (the UN and the OECD). In addition, there is a regulatory pressure of governments all over the world to increase gender diversity. Moreover, between 2009 and 2011, the GMI Ratings' (2012) Women on Boards Survey[f] shows low percentages of women on boards: only 12.9% in Germany, and 12.6% in the USA. In France, the parliament introduced in 2011, a gender quota law to have more gender-balanced boards: the gender quota law of Copé-Zimmermann. French listed firms must appoint at least 20 % of women to their boards by the end of 2010 and at least 40 % by the end of 2017". Consequently, many firms have suddenly increased gender diversity in their boards to comply with this law.

Another dimension of diversity in boards is the director's age. Prior studies suggest that age diversity enhances CSR performance [Ferrero *et al*., 2015; Hafsi and Turgut 2013; Post *et al*., 2011, and Harrison and Klein, 2007]. According to Ouma *et al*. [2017], age diversity could reflect directors' knowledge, experience, and openness to new ideas. Besides, Ferrero-Ferrero *et al*. [2015] argue that age diversity helps to avoid the threat of "narrow group thinking".

Regarding the influence of age diversity on CSR performance, results are mixed. Hafsi and Turgut [2013] and Kets de Vries *et al*. [1984] argue that as directors mature, their generational behavior increases, and therefore, older directors are more likely to be sensitive to society. However, other studies document a negative association between the director's age and CSR performance. For instance, Post *et al*. [2011] argue that younger directors show more concern about environmental issues and tend to be more sensitive to ethical issues.

Also, the presence of foreign directors could be a valuable resource for businesses, specifically on CSR areas [Hafsi and Turgut, 2013; Tihanyi *et al*., 2005; Oxelheim and Randoy, 2003; and Eskeland and Harrison, 2002]. For instance, according to Tihanyi *et al*. [2005] and Eskeland and Harrison [2002], foreign directors are more concerned about philanthropic contributions and local social development. Most often, they have access to broader social networks, diversified and international expertise, and may prefer using technologies producing less waste and pollution. In addition, foreign directors could take advantage of their cultural values on the role of corporations in society to benefit the business [see among others Hafsi and Turgut, 2013 and Oxelheim and Randoy, 2003].

---

[f]    https://www.boardagender.org/files/GMI-Ratings-2012-Women-on-Boards-Survey-F.pdf

Another dimension of diversity is the educational level diversity [Rupley *et al*., 2012; Goll and Rasheed, 2004; Hillman and Dalziel, 2003; and Geletkanycz and Black, 2001]. For instance, Geletkanycz and Black [2001] and Hambrick and Mason [1984] argue that directors with high educational levels contribute to the firm's success, as they have a better capacity to benefit from opportunities and to learn more about new trends. Moreover, Finkelstein *et al*. [2009], Goll and Rasheed [2004]; and Grimm and Smith [1991] suggest that high-educated directors are more likely to adjust their strategies in response to deregulation and other changes, and display different and rational decision-making processes, in comparison with other directors. Furthermore, several studies argue that high-educated directors tend to be more concerned about environmental issues and international markets, to better understand problems that may affect the environment [Shahgholian, 2017; Ewert *et al*., 2001; and Hines *et al*., 1987]. For instance, Shahgholian [2017] put forward that highly educated directors are more likely to help the board to develop environmental activities, as they have more knowledge of environmental issues.

Previous research also shows evidence that multiple directorships could be a key determinant of the involvement in CSR activities [Rupley *et al*., 2012]. In fact, according to Rupley *et al*. [2012], directors who are sitting on multiple boards could bring to the board information about unfamiliar practices, based on their experience on other firms. Therefore, they could help the company to adopt policies of other companies, and increase environmental performance. They are more likely to have access to more information about environmental initiatives and to help to shape more proactive environmental strategies [De Villiers *et al*., 2011]. Having different experiences could increase CSR sensitivity directors with multiple directorships. Accordingly, they could show more ethical behavior and become more involved in CSR practices.

Finally, despite the fact that there is a growing number of studies on diversity in boards and diversity of boards, to the best of our knowledge, the interaction between the two forms of diversity and how it could influence CSR is not yet fully explored.

### II.2 CSR performance and CSR committees

According to Godos-Díez *et al*. [2018], companies establish specialized committees to better deal with a wide range of board functions. For instance, firms concerned about being socially and environmentally responsible, create CSR committees (CSRC) in charge of CSR strategies. Several emerging papers show that companies establish CSRCs to signal their transparency in the field of CSR and their commitment towards sustainable development [Hussain *et al*., 2018a; Mallin and Michelon, 2011; and Eccles *et al*., 2011]. These committees are prone to shape the strategies required to promote and implement firms' CSR initiatives and to decrease CSR risks [Hussain *et al*., 2018a; Peters and Romi, 2015; and Rodrigue *et al*., 2013]. Prior literature shows that the presence of CSRCs on boards increases CSR performance [Cucari *et al*., 2018; Helfaya and Moussa, 2017; Khan, 2017; Peters and Romi, 2015; Walls *et al*., 2012, and Mallin and Michelon, 2011]. For instance, Sánchez *et al*. [2019] and Khan [2017] point out that the existence of CSRCs allows for a better understanding of the key strategic problems facing the board of directors. CSRCs aim to promote CSR strategies and provide new incentives to CEOs to get the business actively involved in CSR projects.

The survey of the literature shows that CSRCs could be also called environmental committees [Liao *et al*., 2015; Walls *et al*., 2012; and Adnan *et al*., 2010]. Environmental committees are established to increase the firm's proactivity in handling environmental issues [Walls *et al*., 2012]. For instance, using a sample of 4,013 firm-year observations from listed companies in 13 European countries from 2002 to 2016, Haque and Jones [2020] find that 63% of the firms maintain CSR committee of the board. Moreover, Eberhardt-Toth [2017] points out that, in 2012, the proportion of firms with CSRCs on boards represents 25.42% in UK, and 16.38% in the USA.

The state of art on CSRC composition shows that most of the current studies have focused on the analysis of structural and demographic characteristics of CSRC members.

### II.2.1 Structural characteristics of CSRC

Turning to the CSRC attributes, to the best of our knowledge, few studies have investigated the CSRC composition. For instance, some papers have analyzed how the presence of independent members in CSRC could affect CSR performance [Danvila del Valle *et al*., 2013; Adams *et al*., 2010]. According to Lovdal *et al*. [1977], in order to maintain a critical view of management operations, 80% of CSRC's directors should be independent.

Also, Danvila del Valle *et al*. [2013], and Aboody and Lev [2000], independent members of CSRCs are more likely to ensure effective monitoring and better management quality, as they provide more objective feedback on firms' activities. Accordingly, their presence could prevent stakeholders from the opportunistic behavior of managers, which could enhance social performance. Another set of papers shows that the presence of independent directors could decrease CSR performance. One explanation is that independent members could suffer from a lack of information about the day-to-day operations of the business and the company's strategies [Adams *et al*., 2010].

They may face limited-access to firm-specific information. Their decisions could be, therefore, largely based on information provided by the managers [De Villiers *et al*., 2011; Donnelly *et al*., 2008; and Adams and Ferreira, 2007].

Another interesting feature of CSRC is CEO membership. In fact, Graham *et al*. [2017] argue that the CEO is prone to care more about profitable projects than environmental and social projects, and could, therefore, avoid risks and uncertainty related to CSR activities. Also, Danvila del Valle *et al*. [2013] point out that it could be difficult to challenge the CEO on CSR issues if he or she is a member of the CSRC, which may affect negatively CSR performance. According to the agency theory, CEOs could adopt an opportunistic behavior to increase their private benefits at the expense of shareholders. Powerful CEOs could be tempted to manipulate information on CSR investments, particularly when they are entrenched [Bebchuk *et al*., 2011; Bartov and Mohanram, 2004; and Aboody and Lev, 2000]. In fact, they could influence committees' discussions, by sharing their personal views in order to maximize their benefits [Clune Richard *et al*., 2014].

Some studies have focused on the relationship between CSRC size and CSR performance. According to Golden *et al*. [2001], significant strategic changes are related to smaller board sizes. This idea can be extended to the CSRC. In fact, strong strategic changes are needed to achieve CSR performance. In a small committee, each director's decision is less likely to depend on the other director's decision. Therefore, directors make more individual effort to fulfill their responsibilities. However, in line with the resource dependence theory [Pfeffer and Salancik, 1978], more directors imply more resources and larger networks, which could be valuable to enhance CSR strategies [Mangena and Pike, 2005; and DeFond and Francis, 2005]. Additionally, Bedard *et al*. [2004] put forward that a larger committee has the necessary strength, diversity of expertise and views to ensure appropriate monitoring' which leads to higher CSR performance.

Not only the CSRC size has been discussed in the literature but also the committee functioning has been analyzed as it could matter in CSR.

For instance, previous studies show that the number of meetings organized could be considered as a proxy for directors' monitoring effort [Nurulyasmin *et al*., 2017; Ponnu et Karthigeyan, 2010; and Vafeas, 1999]. According to Vafeas, [1999], meetings' frequency could be a remedy to asymmetric information problems. For instance, Nurulyasmin *et al*. [2017] and Ponnu and Karthigeyan [2010] put forward that with a high-frequency meeting committee, directors could be more informed about existing and appropriate strategies. Accordingly, they become more likely to use their knowledge to help managers to enhance their decision-making process. A higher frequency of board meetings allows the directors to better carry out their duties in line with shareholders' expectations [see among others Salim *et al*., 2016 and Chou *et al*., 2013].

Another interesting feature of CSRC functioning is directors' assiduity in CSRC meetings. In fact, Huilong *et al*. [2014] argue that directors' assiduity shows the level of commitment to the job, which could have an impact on a firm's corporate governance. This could enhance information sharing between firm management and CSRC, which may increase CSR performance.

### II.2.2 Demographic characteristics

Regarding the demographic characteristics of CSRC, very little evidence has been provided. In fact, the most discussed feature is the presence of female members on CSRC.

If gender diversity is valuable in boardrooms, there is some evidence that it could be also valuable in CSRCs: Appointing more female members on CSRCs on committees could be another way for directors to get actively involved in CSR activities and to effectively enhance social and financial performances [Pucheta-Martínez *et al*., 2016; Krishnan and Parsons, 2008; and Carter *et al*., 2003]. For instance, Pucheta-Martínez *et al*. [2016] and Carter *et al*. [2003] point out that female directors are more prone to improve the performance of the CSRC, by increasing the creativity and innovation of the committee. In fact, they could bring important resources such as skills and constituencies and external networks [Krishnan and Parsons, 2008].

Besides, previous studies also show that the significant roles of boards are usually attributed to the more interactive and participative style of leadership by female directors [Conyon and He, 2017; and Elstad and Ladegard, 2012]. The firms rely more on female directors' skills, especially when the chair committee is a woman [Peterson and Philpot, 2007; and Mattis, 2000].

To the best of our knowledge, there is a huge gap in the literature on the influence of other demographic features on CSRC such as age, the educational level, the professional experience and ethnic diversity on social performance.

### (III) Discussion

The current chapter sheds light on urgent and critical issues on boards and their functioning when it comes to CSR-related concerns.

From a managerial perspective, it shows the relevance of the board composition in CSR strategies. Boards have to prioritize structural and demographic diversities to bring new meaningful insights specifically in terms of more ethical behavior. Increasing diversity could boost overall corporate visibility and develop a more proactive and comprehensive CSR strategy and orientation.

First, the emerging studies on the Board-CSR relationship show how the diversity of boards and diversity in boards could help the firm to achieve a double-target: improving the governance quality and getting better social and financial performances.

In practice, there is a huge debate on the way to achieve more diverse-balanced boards. Some studies stand for the role of regulation to increase diversity in boards (like for example, gender quota law). However, this could lead to the increase of multiple directorships and raises, therefore, several questions on the extent to which directors sitting on different boards could be and stay independent.

For instance, in France and Spain, as the pool of business women candidates and the need to urgently comply with the law, many firms have appointed non-business women members to their boards. Some non-regulation defenders highlight the side effects of appointing non-expert profiles and its influence on the business. Accordingly, as a response, many countries, have preferred to rely on recommendations and voluntary quotas, instead of laws to increase the involvement of minorities in top management positions.

Besides, when minority groups are not supported in their choices, this could worsen the board dynamics, and slower therefore, the decision-making process. For instance, Erkut *et al*. [2008] show that when only one or two female members are appointed to the board, they cannot influence the boards' decisions. A critical number of female board members is needed in order to exert a positive influence on the boards' decisions.

Furthermore, women are more likely than men to serve in precarious management positions and continue to be underrepresented in leadership positions. This is explained by the "glass ceiling" that prevents women from gaining access to such positions [Singh and Vinnicombe, 2004; Arfken *et al*., 2004; and Maume, 2004] and the "glass escalator", which means that men are accelerated through the organizational ranks [Maume, 1999; and Williams, 1992].

Also, the current chapter shows that most of the existing studies on Board-CSR relationship are more likely to explore one-direction of this association, according to which governance features, and therefore CSR committees, help to achieve higher social and financial performances. However, the question of the potential reversal effect of financial performance on governance quality, more specifically on the effectiveness of CSR committees and their compositions, are not yet analyzed. To the best of our knowledge, no studies have examined how financial performance could influence board composition and functioning.

Finally, identifying the timely role of CSR committees in improving CSR strategies should not be taken without consideration of the role of other board committees specifically nomination committees (NC) that are in charge of selecting new board members candidates. NC have definitely influential effects on diversity in (and of) boards and therefore on CSRC composition. To the best of our knowledge, studies on NC role and composition are still scarce.

Shedding more light on the cognitive and individual characteristics of committee members is valuable to understand the board dynamics and how they influence firms' strategies (see figure 1.1).

Fig. 1.1. Theoretical model

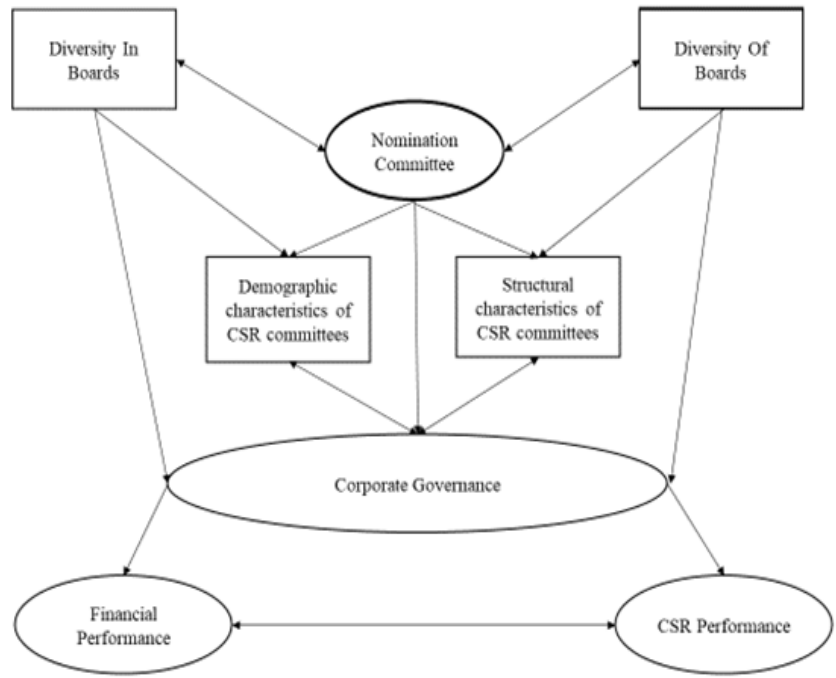

**Conclusion**

Companies have become more concerned about the protection of human rights, employees' conditions, environmental issues, and communities' expectations. They manage their business according to specific ethical standards. Enhancing governance quality is also among the challenging issues in CSR. The main aim of the current chapter is to put forward the influence of governance features on CSR. It covers these challenging issues.

First, the results of recent empirical studies remain very mixed and the majority of these studies show the existence of a positive association between social performance and financial performance [Rost and Ehrmann, 2017; Endrikat *et al*., 2014; Albertini, 2013; Dixon-Fowler *et al*., 2013; Margolis *et al*., 2009; and Orlitzky and Swanson, 2008]. This could be explained by the difference in terms of CSR strategies adopted by firms.

Second, previous studies show that strategic CSR is a source of original and pioneering actions, where interactions with stakeholders are the key to sustainable performance. Strategic CSR aims to create resources and capabilities that can lead to superior economic performance. While responsive CSR reflects more adaptive behavior. Companies try to gain legitimacy in the eyes of the firm's stakeholders to appear socially responsible [Ruggiero *et al*., 2018].

Third, there is an emerging literature on how board diversity could be an advantage for the decision-making process and the key to strategic CSR [Bocquet *et al*., 2019]. This heterogeneity can promote diversified exchanges and relationships, offer new perspectives, and influence the board's functioning, which in turn can influence its performance [see among others Isidro and Sobral, 2015; and Aggarwal and Dow, 2012]. Specifically, taking into account structural characteristics of boards such as size, duality structure, and board independence, and director's profile such as gender, age, foreign directors and educational level, could be very helpful to better understand how boards of directors influence firm performance from different perspectives [Haniffa and Cooke, 2005; and Gibbins *et al*., 1990].

Finally, CSR committees have attracted increasing interest [Khan, 2017; Peters and Romi, 2015; and Rodrigue *et al*., 2013]. Specifically, previous studies show that the presence of CSR committees acts as an effective mechanism to enhance CSR performance [Khan, 2017]. They could help to promote and implement firms' CSR initiatives, which decreases CSR risks and achieve new opportunities [Perters and Romi, 2015; and Rodrigue *et al*., 2013].

In future research, it could be interesting to focus on the financial and strategic risk-taking in a responsive or strategic form of CSR.

**Biography of the authors**

Author 1: Rania Beji
o PhD Student in Finance, Research Assistant at Montpellier Business School
o Member of MRM Lab, Université de Montpellier (France) and GEF-2A Lab, Institut Supérieur de Gestion de Tunis, Université de Tunis (Tunisia)

Author 2: Ouidad Yousfi
o Associate Professor of Finance at IUT Montpellier
o Member of MRM Lab, Université de Montpellier (France)

Author 3: Abdelwahed Omri
o Professor in Financial and Accounting Methods at Institut Supérieur de Gestion de Tunis
o Director of GEF-2A Lab, Université de Tunis (Tunisia)